\documentclass[twocolumn,showpacs,amsmath,amssymb]{revtex4}

\usepackage[dvips]{graphicx}
\usepackage[dvips]{color}

\begin{document}
\definecolor{Red}{rgb}{1,0,0}



\title{Finite genome length corrections for the mean fitness
and gene probabilities in evolution models}
\author{Zara Kirakosyan$^{1}$}
\author{David B. Saakian$^{1,2,3}$}\email{saakian@phys.sinica.edu.tw}
\author{Chin-Kun Hu$^{2,4}$}
\email{hu@phys.sinica.edu.tw}
 \affiliation{$^1$Yerevan Physics
Institute, Alikhanian Brothers St. 2, Yerevan 375036, Armenia}
\affiliation{$^2$Institute of Physics, Academia Sinica, Nankang,
Taipei 11529, Taiwan}

\affiliation{$^3$National Center for Theoretical Sciences:Physics Division,
National Taiwan University, Taipei 10617, Taiwan}

\affiliation{$^4$Center for Nonlinear and Complex Systems and Department of Physics,
  Chung Yuan Christian University, Chungli 32023, Taiwan}

\date{\today}

\begin{abstract}

Using the Hamilton-Jacobi equation approach to study genomes of
length $L$, we obtain $1/L$ corrections for the steady state
population distributions and mean fitness functions for horizontal
gene transfer model, as well as for the diploid evolution model
with general fitness landscapes. Our numerical solutions confirm
the obtained analytic equations. Our method could be applied to
the general case of nonlinear Markov models.

\end{abstract}


\maketitle


\section{Introduction}


After the development of molecular biology, the concepts and
methods in statistical physics have been widely used to study
molecular models of biological evolution
\cite{ei71,ei89,ss82,ck70,le87,ta92,bar92,ba95,ko99,fp97,bb97,ba01,ba02},
especially in recent years
\cite{sa04,sa04a,le05,sa06,sa06a,de06,de07,sa07,sa08,de08,sa10}.
Analytic solutions for models with the infinite genome length may
be obtained via several methods, including the maximum principle
\cite{ba01,ba02}, Trotter-Suzuki method in quantum statistical
physics \cite{sa04,sa04a,sa06,sa06a}, quantum field theory
\cite{sa04a,sa06a,de06} and the Hamiltonian-Jacobi equation method
\cite{sa07,ka07,sa08,sa10}. The infinite genome length limit has
been solved fora  more complicated (more nonlinear) evolution
model: the horizontal gene transfer model
\cite{le05,de07,sa08,sa10}. This model has been solved both for
the haploid \cite{le05,de07}, and the hyper-geometric diploid case
\cite{sa08,bar92,ba95,ko99}.

The study of finite-size corrections is important not only for
lattice phase transition models, but also for population genetics
\cite{ew04,bu00}. Here there are situations with few relevant
genes for the evolution problem. Although the genome length may be
large, e.g. about 10000 in the case of HIV, it is possible to
consider only the variable part of the genome which has only
$40-100$ bases \cite{bo05}; in such a case, the finite-size
effects cannot be overlooked. The study of finite-size corrections
allows us to understand as to what extent the results for the
infinite genome length can be used to represent those for the
finite genome length.

In recent works \cite{de06,sa07} the finite genome length
corrections have been calculated for the haploid molecular
evolution model \cite{ck70}. In \cite{sa10} we calculated the
finite size correction for the recombination model with
single-peak fitness. In the present paper we calculate the finite
genome length corrections for a diploid model with symmetric
landscape \cite{sa08} as well as for a haploid model with a simple
horizontal gene transfer (HGT) \cite{le05,de07} for a general
symmetric fitness landscape. The method may be applied to rather
general cases of nonlinear probabilistic models \cite{fr04}.

Actually, we are constructing a perturbation expansion for the
eigen-value like variable for the nonlinear operators.
 In case of linear operators (like the quantum mechanics),
to calculate the first order perturbation to the energy we need
only  the 0-th order expressions of eigen-vectors to calculate the
leading corrections to eigen-values due to perturbations. In case
of nonlinear system of equations it is impossible to use the
methods of linear algebra. Nevertheless, we can succeed using a
trick. To calculate the perturbation  expansion of eigen-value
like variables (mean fitness or the asymptotic expression), we use
equations  such as Eq.(13), where the contribution of correction
terms of the distribution function disappear. Thus the
perturbation of mean-fitness ("eigen-value" for a nonlinear
problem ) can be calculated using only the zeroth-order expression
of eigen-functions (bulk solutions for probability distribution).


The present paper is organized as follows. In Section II we derive
the finite size corrections for a haploid model (Crow-Kimura
model) with a new method. In Sections III and IV the new method is
applied to the diploid evolution model and HGT model,
respectively. In the Appendix we give finite size corrections for
the diploid evolution model with single-peak fitness function.

\section{Finite size corrections for the Crow-Kimura model}


Let us first introduce a new method to calculate finite-size
corrections for the Crow-Kimura model with haploid genotypes
\cite{ck70}, which is easier to study. The Crow-Kimura  model is
slightly easier to solve than the Eigen model \cite{ei71,ei89}.


 In the Crow-Kimura model
\cite{ck70}, any genotype configuration
${S_i}\equiv(s^i_1,....,s^i_L)$ is specified by a sequence of
\textit{L} two-valued spins $S^i_k=\pm 1$ for $L \ge k \ge 1$ and
$M-1 \ge i \ge 0$ such that the value $+1$ represents purines
(A,G), and the value $-1$ represents pyrimidines (T,C), where
$M=2^L$ is the total number of different sequences. The difference
between two configuration ${S_i}$ and $S_j$ is described by the
Hamming distance
$d_{ij}=(L-\sum_k s_k^is_k^j)/2$.
In other words, the Hamming distance is the number of different
spins between configurations $S_i$ and $S_j$.

The relative frequency $x_i$ of a given configuration $S_i$
satisfies following equation:
\begin{eqnarray} \label{e1}
\frac
{dx_i}{dt}=x_i(r_i-\sum_{j=0}^{M-1}r_jx_j)+\sum_{j=0}^{M-1}\mu_{ij}x_j,
\end{eqnarray}
 here $r_i$
is the replication rate, i.e. the \textit{fitness} of an organism
with a given genotype and in Crow-Kimura's model it is specified
like a function of the genotype: $r_i=Lf(S_i)$. In other words
\textit{fitness} is the average number of offspring of individual
with genotype sequence $S_i$ per unit period of time and it is
very meaningful quantity in evolution theory. $\mu_{ij}$ is the
mutation rate to move from sequence $S_i$ to sequence $S_j$ per
unit period of time. It is important that in Crow-Kimura's model
\cite{ck70} only single base mutations are allowed, i. e.
$\mu_{ij}=\mu$ for $d_{ij}=1$, $\mu_{ij}=0$ for $d_{ij}\ne 1$ and
$i \ne j$, and $\mu_{ij}=-L\mu$ for $i=j$; the last condition
ensures that the time evolution of $x_i$ does not change the
normalization condition $\sum_{j=0}^{M-1} x_j=1$. In the
following, we take $\mu=1$.

The nonlinear system of Eq. (\ref{e1}) can be mapped into
following linear system of equations,
\begin{eqnarray} \label{e2}
\frac {dp_i}{dt}=p_ir_i+\sum_{j=0}^{M-1}\mu_{ij}p_j,
\end{eqnarray}
where $p_i$ are related to $x_i$ of Eq.(\ref{e1}) via
\cite{tb74,jer75},
\begin{eqnarray} \label{e3}
 x_i=\frac{p_i}{\sum_jp_j}.
\end{eqnarray}

For the symmetric fitness landscape, where $r_i$ are the same for
all the sequences with the same Hamming distance from the same
reference sequence $S_0\equiv (1,1,\dots,1)$, it is more
convenient to work with \textit{classes}, i.e. we classify
configurations $S_i$ into the classes according to the value $m_l
\equiv m_i$, where $m_i$ is so-called ``magnetization'' of the
configuration $S_i$ and is defined as $m_i=\sum_{k=1}^L s_k^i/L,
-1 \leq m^i\leq 1$.

 Defining the magnetization
$m_l$ for the configurations at the $l$-th class as $m_l=1-2l/L$
and fitness $r_i\equiv Lf(m_l)$, we rewrite  Eq.(\ref{e2}) for the
probability $p_l$ for a typical configuration in the $l$-class as
\begin{eqnarray} \label{e4}
\frac{dp_l}{dt}=p_l(L f(m_l)-L)+lp_{l-1}+(L-l)p_{l+1}.
\end{eqnarray}
We define the $l$-th Hamming class as the group of all sequences at
the Hamming distance $l$ from the reference sequences.
 Having the number of configurations at the $l$-th class
$L_l=\frac{L!}{l!(L-l)!}$ for $P_l=L_lp_l$ we have
\begin{eqnarray} \label{e5}
\frac{dP_l}{dt}&=&P_l(L f(m_l)-L)+(L-l+1)P_{l-1} \nonumber \\
&+&(l+1)P_{l+1}.
\end{eqnarray}

 We are interested in calculating the mean fitness $R\equiv \sum_lP_lf(m_l)$ and surplus
$s\equiv \sum_lP_lm_l$ in
  the steady state.
In \cite{sa07}, one of us proposed that in the steady state $\ln
P_l$ and the mean fitness $R$ can be written as
\begin{eqnarray} \label{e6}
\ln P_l&=&L U(m,t)+O(1),\nonumber\\
R&=&Lk+O(1),
\end{eqnarray}
where $m=1-2l/L$. Then, high order corrections for $R$, and for
$\ln P_l$ have been derived using the linear algebra methods and
the equations for $P_l/\sqrt{L_l}$. In this section we derive
these corrections with an alternative method that can be applied
to the strong nonlinear situations.

 Using the ansatz
$P_l=exp[LU(m,t)]$ and a formula $P_{l\pm 1}\approx P_l e^{-(\pm
2U'_m)}$ where $U'\equiv \frac{\partial U(m,t)}{\partial m}$, one
can transform  Eq. (\ref{e5}) into the Hamilton-Jacobi equation,
\begin{eqnarray} \label{e7}
\frac{\partial U}{\partial
t}&=&H(m,U'), \nonumber\\
H(m,y)&\equiv& f(m)-1+\frac{1+m}{2}e^{2y}+\frac{1-m}{2}e^{-2y},
\end{eqnarray}
where $y \equiv U'$ is a dummy variable. We assume an asymptotic
\begin{eqnarray} \label{e7a}
U(m,t)=kt+u(m),
\end{eqnarray}
 and get an ordinary  differential equation
\begin{eqnarray}
 \label{e8}
k=f(m)-1+\frac{1+m}{2}e^{2u'}+\frac{1-m}{2}e^{-2u'}\nonumber\\
2u'=\ln \frac{k-1+f(m)\pm \sqrt{(k-1+f(m))^2-1+m^2}}{1+m},
\end{eqnarray}
which corresponds to Eq. (23) in Ref. \cite{sa07}. We take the "+"
solution for $-1\le m<m_0$ and $"-"$ solution for $m>m_0$.

 In Eq.
(\ref{e8}), $k$ is a function $u'$. The value of $u'$, which gives
the maximum value of $k$ is defined by equation
\begin{eqnarray}
 \label{e8b}
u'_0(m)=\frac{1}{4}\ln \frac{1-m}{1+m}.
\end{eqnarray}
Thus the maximum value of $k$ can be obtained from
\begin{eqnarray}
 \label{e8a}
k&=&max[V(m)]_{-1\le m\le 1}, \nonumber\\
V(m)&\equiv& H(m,u_0(m))=f(m)-1+\sqrt{1-m^2},
\end{eqnarray}
where the maximum of the first equation is at the point $m_0$.

To calculate higher order corrections, we write $\ln P_l$ as
\begin{eqnarray} \label{e9} \ln
P_l=L(k+\frac{k_1}{L})t+Lu(m)+u_1(m)+O(1).
\end{eqnarray}
Equations (\ref{e5}) and (\ref{e9}) imply that \cite{sa07}:
\begin{eqnarray}
\label{e10} &&k+\frac{k_1}{L}=
f(m)-1+\frac{1+m}{2}e^{2u'}+\frac{1-m}{2}e^{-2u'} \nonumber \\
&+&\frac{1}{L}(e^{2u'}+e^{-2u'})+\frac{2u''}{L}(\frac{1+m}{2}e^{2u'}+\frac{1-m}{2}e^{-2u'})\nonumber\\
&+&\frac{2u_1'}{L}[\frac{1+m}{2}e^{2u'}-\frac{1-m}{2}e^{-2u'}].
\end{eqnarray}
In Eq.(\ref{e10}), $k$ is the bulk expression of the mean fitness
and $k_1/L$ is the first order correction to it. Having the value
of $k_1$, we can calculate $u_1'$.  $k_1$ can be defined from
 Eq.(\ref{e10}) at the point $m=m_0$, where the
coefficient of $u_1'$ becomes zero. We have an equation
\begin{eqnarray}
\label{e11} e^{2u'(m_0)}=\sqrt{\frac{1-m_0}{1+m_0}}.
\end{eqnarray}

Equation (\ref{e10}) then implies that at $m_0$
\begin{eqnarray}
\label{e13} k_1&=&
e^{2u'}+e^{-2u'}+{2u''}(\frac{1+m}{2}e^{2u'}+\frac{1-m}{2}e^{-2u'})\nonumber\\
&=&2\frac{1}{\sqrt{1-m_0^2}}+2u''(m_0)\sqrt{1-m_0^2}.
\end{eqnarray}
Let us define $u''(m_0)$. Equation (\ref{e8}) can be written as
$k=H(m,p=u')$.  Expanding Eq. (\ref{e8}) near $m=m_0$ with respect
to the first and the second arguments of $H(m,p=u')$ up to the
second order, we derive

\begin{eqnarray}
\label{e13a} 0 &\approx& V''(m_0)\frac{(m-m_0)^2}{2} \nonumber \\
&+&H''_{pp}(m_0,u_0(m_0))\frac{(u'(m)-u'_0(m))^2}{2}.
\end{eqnarray}
Note that there is no $(m-m_0)(u'-u'_0)$ term in the last
equation. We can verify this directly: $H'_p(m,p_0)=0$.

From Eqs. (\ref{e7}), (\ref{e8a}) and (\ref{e8b}), we  derive
$$H''_{pp}=4 \sqrt{1-m_0^2},$$
$$V''(m_0)=f''(m_0)-\frac{1}{(1-m_0^2)^{3/2}},$$
$$u'_0(m)\approx -(m-m_0)\frac{1}{2(1-m_0^2)}+u'_0(m_0).$$
With these expressions in Eq.(\ref{e13a}), we derive
$$f''(m_0)=\frac{1}{(1-m_0^2)^{3/2}}-
\sqrt{1-m_0^2}(2u''(m_0)+\frac{1}{1-m_0^2})^2.$$ Therefore
\begin{eqnarray}
\label{e14}
2u''(m_0)&=& -\frac{1}{(1-m_0^2)}                \nonumber\\
&-&\frac{1}{(1-m_0^2)^{1/4}}\sqrt{f''-\frac{1}{(1-m_0^2)^{3/2}}}.
\end{eqnarray}

Equations (\ref{e8b}) and (\ref{e11})  imply that
 $u'(m_0)=u'_0(m_0)$. Thus eventually, we can use
Eq. (\ref{e13}) to obtain \cite{de06},\cite{sa07}
\begin{eqnarray}
\label{e15} &&k_1= \frac{1}{\sqrt{1-m_0^2}} \nonumber \\
&&-\sqrt{\frac{1}{(1-m_0^2)^{3/2}}-f''(m_0)}(1-m_0^2)^{1/4}\nonumber\\
&=& \frac{1}{\sqrt{1-m_0^2}}[1-\sqrt{1-(1-m_0^2)^{3/2}f''(m_0)}].
\end{eqnarray}
Using the expression for $k_1$, we can calculate the $u_1'$ from Eq.
(\ref{e10}):
\begin{eqnarray}
\label{e16} &&2 u_1'[\frac{1+m}{2}e^{2u'}-\frac{1-m}{2}e^{-2u'}]
\nonumber\\
&=&k_1-(e^{2u'}+e^{-2u'}) \nonumber \\
&-&2u''(\frac{1+m}{2}e^{2u'}+\frac{1-m}{2}e^{-2u'}).
\end{eqnarray}
Therefore, having $u_1$, Eq.(\ref{e16}), we can calculate the
$1/L$ accuracy expression for the haploid class probabilities
$P_l$ (we define them as $\hat{p_{1l}}\equiv(Lu(m)+u_1(m))$), also
we can calculate the probabilities $p_{0l} \equiv e^{L u(m)}$,
Eq.(\ref{e9}). The comparison of this computations with numerical
results are shown in the Fig. 1 for the case $L=80,160$.

\begin{figure} \large \unitlength=0.1in
\begin{picture}(42,12)
\put(-1.7,0){\includegraphics{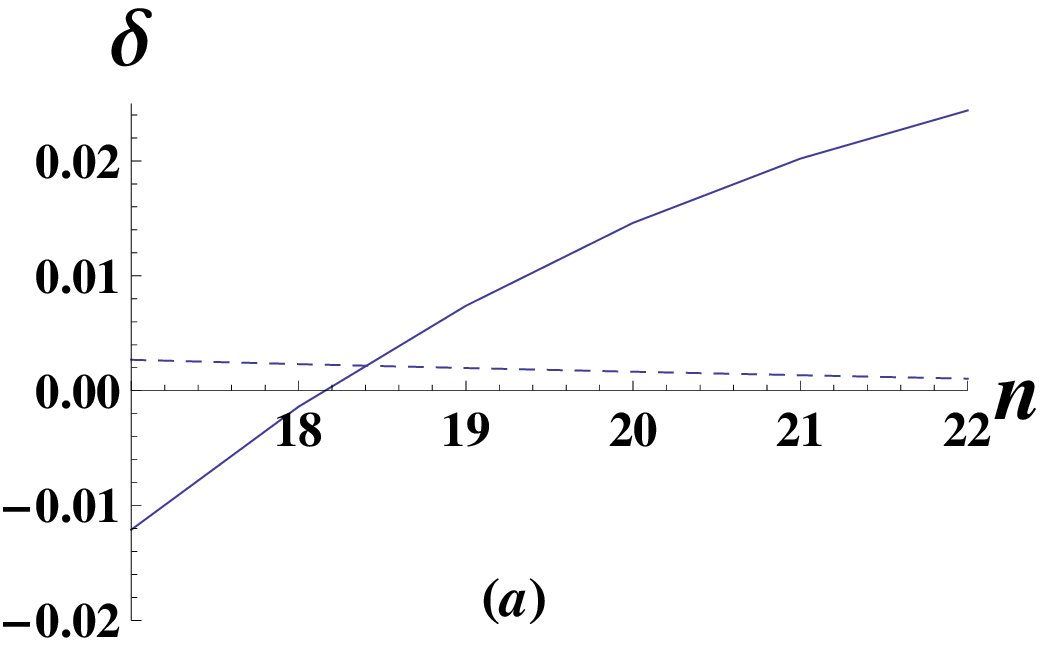}}
\put(16.5,0){\includegraphics{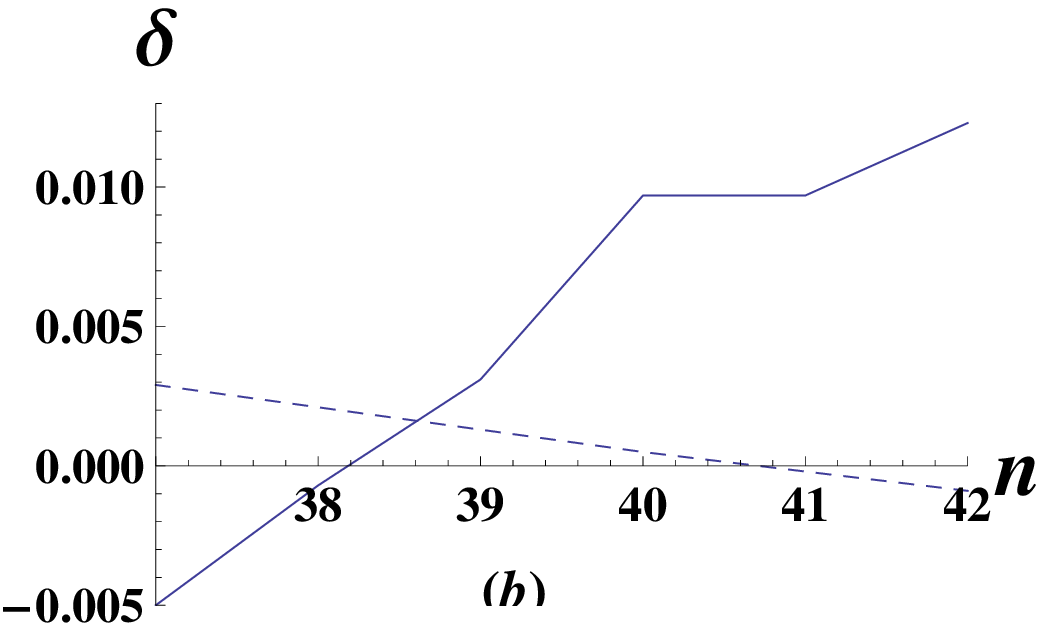}}
\put(7.7,0){\small{a.}} \put(26,0){\small{b.}}
\end{picture}
\caption{
 Comparison of the 0-th order and $1/L$ accuracy
expressions for the haploid model class probabilities with the
numerics given by Eq.(4) for the fitness function
$f(m)=\frac{1}{2}m^2$.
 $p_n$ are the
results of numerics for the original model given by the system of
equations. $p_{0n}\equiv\exp(Lu(m))$ is calculated using $O(L)$
terms in expression of $\ln p_n$, Eq.(12), while
$p_{1n}\equiv\exp(Lu(m)+u_1(m))$ corresponds to the $O(1)$
accuracy in expression of $\ln p_n$, Eq. (12). The smooth line
corresponds to $\delta=\ln p_{0n}/p_n$. For Fig 1.a $ L=80$ and
the dashed line corresponds to $\delta=\ln p_{1n}/p_n$. For Fig
1.b $L=160$ and the dashed line corresponds to $\delta=10\ln
p_{1n}/p_n$.}
\end{figure}

On the other hand, we can calculate $k_1$ directly as
$\sum_lf(m_l)P_l-k$,
\begin{eqnarray}
\label{e18} k_1=\frac{1}{|u''(s)|}(f'(s)u_1'(s)+\frac{f''(s)}{2}).
\end{eqnarray}
The bulk  surplus term is calculated using the equation
\cite{ba01}
$$f(s)=R.$$
From the definition of $s$, the correction to the surplus can be
written as $\delta s=\frac{u_1'(s)}{L|u''(s)|}$. Comparing with
Eq. (\ref{e18}), we get $\delta
s=(k_1-\frac{f''(s)}{2|u''(s)|})/(Lf'(s))$. Using the formula
\begin{eqnarray}
\label{e19}u''(s)=-\frac{f'(s)}{2s},
\end{eqnarray}
we derive, assuming $f'(s)>0$:
\begin{eqnarray}
\label{e20}\delta s=(k_1-\frac{sf''(s)}{f'(s)})\frac{1}{Lf'(s)}.
\end{eqnarray}

One can define $u_1(m)$ integrating the expression of $u'_1(m)$
from Eq. (\ref{e16}). Thus we have derived the results of
\cite{sa07} by an alternative method.

\section{First order corrections for the diploid case}

The diploid model has been defined by Eq. (1) in \cite{sa08}.
There we derived the analytic  solution for the diploid many
allele biological evolution models \cite{bar92,ba95,ko99} with
general fitness landscapes to the first order using the
Hamilton-Jacobi equation approach. In the present paper we will
find $1/L$ accuracy for mean fitness and genome probabilities of
that diploid model.

 In WBS parallel diploid model \cite{ba95},
 gene probabilities $p_i$ evolve  as
\begin{equation}
\label{e21a}
\frac{dp_i}{dt}= p_i[\sum_{j=1}^M A_{ij}p_j
-\sum_{k=1}^{M}\sum_{j=1}^{M}A_{jk}p_jp_k]+\sum_{j=1}^M m_{ij}p_j.
\end{equation}
Here $m_{ij}$ is a mutation matrix, defined in Section II. We have
a balance condition $\sum_{i=1}^{M}p_i=1$. $A_{ij}$ is the fitness
of the genotype $(S_i,S_j)$, and $\sum_j A_{ij}p_j$ is the
marginal fitness for the sequence $S_i$.

We assume that the fitness of the configuration $S_i$ is a smooth
function of the Hamming distance between $S_i$ and the reference
configuration $S_1$. In such case, it is convenient to work with
the overlap $m=(1-2d_{i1}/L)$ instead of the Hamming distance
$d_{i1}$. Consider the following choice of the matrix $\hat A$:
\begin{eqnarray}
\label{e21b} A_{ii}=&f(m,m), \nonumber\\
A_{ij}=&f(m_1,m_2),
\end{eqnarray}
where $m_1=1-2d_{i1}/L,m_2=1-2d_{j1}/L$. In the hypergeometric
model $L(1+m_1)/2$ and $L(1+m_2)/2$ are, respectively, the number
of $A_1$ maternal and paternal alleles. $f(m_1,m_2)$ is a smooth
analytical function.
 We are interested in finding the exact phase structure and the steady state,
 therefore we can consider only symmetric solutions of $p_i$.

Using the expression $P_l\sim exp[Rt+\dots]$ in Eq. (13) of
\cite{sa08}, we obtain coupled equations for $P_l$ that describe
the diploid, one locus many allele parallel mutation-selection
model \cite{sa08,note}:
\begin{eqnarray}
\label{e21}
RP_l&=& LP_lF_l+(L-l+1)P_{l-1}+(l+1)P_{l+1} \nonumber\\
&-& LP_l(1+\sum_{k}F_kP_k).
\end{eqnarray}
Here we define the marginal fitness for the Hamming class
$F_l=\sum_{n=0}^L f(1-2l/L,1-2n/L)P_n$. In diploid case the
fitness landscape is defined via a function $f(m_1,m_2)$ of two
arguments, describing the dominance relations among alleles
\cite{bar92,ba95,ko99}. Eq.(21) coincides with the Crow-Kimura
model where $F_l$ is the fitness function for the Hamming class l.

Assuming ansatz (\ref{e6}) and (\ref{e7a}), we derive the bulk
expression for steady state solution as
\begin{eqnarray}
\label{e22} k=
f(m,s)-1+\frac{1+m}{2}e^{2u'(m)}+\frac{1-m}{2}e^{-2u'(m)},
\end{eqnarray}
where $u(m)$ is defined via Eqs.(6),(8). The mean fitness per spin
$k$ and surplus $s\equiv \sum_lP_l(1-2l/L)$ are defined via a system
\cite{sa08}
\begin{eqnarray}
\label{e23} k&=& f(s,s),\nonumber\\
f'_m(m_0,s)&-&\frac{m_0}{\sqrt{1-m_0^2}}=0.
\end{eqnarray}

To find an expression for $F_l$ with $1/L$ accuracy, we calculate
the following integral:
\begin{eqnarray}
\label{e24} F_l=\int f(m,x)e^{Lu(x)+u_1(x)}dx,
\end{eqnarray}
where $m=1-2l/L$.  Therefore we derive
\begin{eqnarray}
\label{e25} F_l-f(m,s)&=&\frac{f'_b(m,s)[u_1'(s)]}{L|u''(s)|}
+\frac{f''_{bb}[m,s]}{2L|u''(s)|} \nonumber\\
&\equiv& f'_b(m,s)\frac{A}{L}+\frac{f''_{bb}[m,s]}{2L|u''(s)|},
\end{eqnarray}
where $A={[u_1'(s)]}/{|u''(s)|}$, $s$ is the surplus, $f'_b(m,s)$
is the first derivative of the fitness function via the second
argument, and $f''_{bb}(m,s)$ is the second derivative of the
fitness function via the second argument.

For the mean fitness $R/L\equiv k+\frac{k_1}{L}$, from Eq.
(\ref{e25}) we have
\begin{eqnarray}
\label{e26} \frac{R}{L} &=&\sum_l F_l P_l=\int
f(m,s)e^{Lu(m)+u_1(m)}dm\nonumber \\
&+&f'_b(m,s)\frac{A}{L}+\frac{f''_{bb}[m,s]}{2|u''(s)|}.
\end{eqnarray}
Expanding $f(m,s)$ via the first argument in the equation above, we
obtain
\begin{eqnarray}
\label{e27}k_1=
[f'_b(s,s)+f'_a(s,s)]\frac{A}{L}+\frac{[f''_{bb}[s,s]+f''_{aa}[s,s]}{2L|u''(s)|}
\end{eqnarray}
with  $f'_a(m,s)$ being the first derivative of the fitness
function via the first argument and $f'_{aa}(m,s)$ being the
second derivative of the fitness function via the first argument.
On the other hand, we can get an expression for $k_1$ by adding
the term $f'_b(m,s)\frac{A}{L}+\frac{f''_{bb}[m,s]}{2|u''(s)|}$ to
the right-hand side of Eq.(\ref{e16}). Thus we obtain the
following equation for the corrections:
\begin{eqnarray}
\label{e28}
&&[f'_b(s,s)+f'_a(s,s)]A+\frac{[f''_{bb}[s,s]+f''_{aa}[s,s]}{2|u''(s)|}\nonumber\\
&=&2u'_1[\frac{1+m}{2}e^{2u'}-\frac{1-m}{2}e^{-2u'}] \nonumber\\
&+&f'_b(m,s)A+\frac{f''_{bb}[m,s]}{2|u''(s)|}\nonumber\\
&+&2u''[\frac{1+m}{2}e^{2u'}+\frac{1-m}{2}e^{-2u'}]+2\cosh(2u').
\end{eqnarray}
The first line is derived via direct integration of the fitness
function via the steady state distribution, the $u_1'$ term in the
second line is just the small $u_1'$ correction of the right hand
side of Eq.(\ref{e22}). The third line corresponds to
Eq.(\ref{e25}). The last line corresponds to the correction terms
of the haploid model.

To calculate $k_1$ we again consider the point where the coefficient
of $u_1'$ disappears in the last equation. We calculate $u''(m_0)$
as in the previous section. We have an equation for $A$
\begin{eqnarray}
\label{e29}
&&[f'_b(s,s)+f'_a(s,s)-f'_b(m_0,s]A\nonumber\\
&=&\frac{f''_{bb}(m_0,s)-f''_{aa}(s,s)-f''_{bb}(s,s)]}{2|u''(s)|}\nonumber\\
&+&\frac{1}{\sqrt{1-m_0^2}}[1-\sqrt{1-f''_{aa}(m_0,s)(1-m_0^2)^{3/2}}]\nonumber\\
&+&\frac{f''_{bb}(m_0,s)-f''_{aa}(s,s)-f''_{bb}(s,s)]}{2|u''(s)|},
\end{eqnarray}
where  $m_0$ is the bulk magnetization. Then putting the value of
$A$ in Eq.(\ref{e29}), we get an expression for $k_1$:

\begin{eqnarray}
\label{e30}&& k_1= \frac{f''_{aa}(s,s)+f''_{bb}(s,s)}{2|u''(s)|}
\nonumber \\
 &+&
 \frac{f'_b(s,s)+f'_a(s,s)}{f'_b(s,s)+f'_a(s,s)-f'_b(m_0,s)}\nonumber
 \\
&\times&[\frac{f''_{bb}(m_0,s)-f''_{aa}(s,s)-f''_{bb}(s,s)}{2|u''(s)|}+k_{1h}],
\end{eqnarray}
$$k_{1h}=\frac{1}{\sqrt{1-m_0^2}}[1-\sqrt{1-f''_{aa}(m_0,s)(1-m_0^2)^{3/2}}],$$
where
\begin{eqnarray}
\label{e31} u''(s)=-\frac{f'_a(s,s)}{2s}.
\end{eqnarray}

From Eq.(\ref{e23}) we calculate first order accuracy expression
for the diploid mean fitness $k_{theor}\equiv k_0$ and then from
Eq.(\ref{e30}) we compute the $1/L$ accuracy expression for mean
fitness $k=k_0+k_1/L$. Having all these expressions, we compare
our analytical results with the direct numerics  for different
values of $L$ and different values of $a$ in Table I.

 Having $u_1$ by
Eq.(\ref{e28}), we calculate the $1/L$ accuracy expression for the
diploid class probabilities $p_{2n}=\exp(Lu(m)+u_1(m))$. In Fig. 2
we give the comparison of our analytical results with the direct
numerics.
\begin{table}[t]
\begin{tabular}{|c|c|c|c|c|c|c|}
\hline $L$               &100 &100&100&150&150&150 \\
\hline $a$               &4.5 &5.5&6&4.5&5.5&6 \\
\hline $k_{n}$ &3.18506 &4.15619 &4.64491 &3.18455 &4.15574 &4.64449\\
\hline $\delta_0'$ &0.00152&0.00132&0.00124&0.001014&0.000883&0.000829\\
\hline $\delta_1'$ &0.00001&0.00001&0.00001&0.000005&0.000005&0.000005\\
\hline
\end{tabular}
\caption{Comparison of 0-th order accuracy expression
$\delta'_0\equiv k_{n}-k_0$ and the 1/L accuracy expression
$\delta'_1\equiv k_{n}-k$ results for the mean fitness of diploid
model, for the fitness function
$f_0(m_1,m_2)=\frac{a}{2}(m_1^2+m_2^2)+ bm_1m_2$,for $b=0.5$.
$k_n$ is given by direct numerics of Eq.(4).}
\end{table}


\begin{figure} \large \unitlength=0.1in
\begin{picture}(42,12)
\put(-1.7,0){\includegraphics{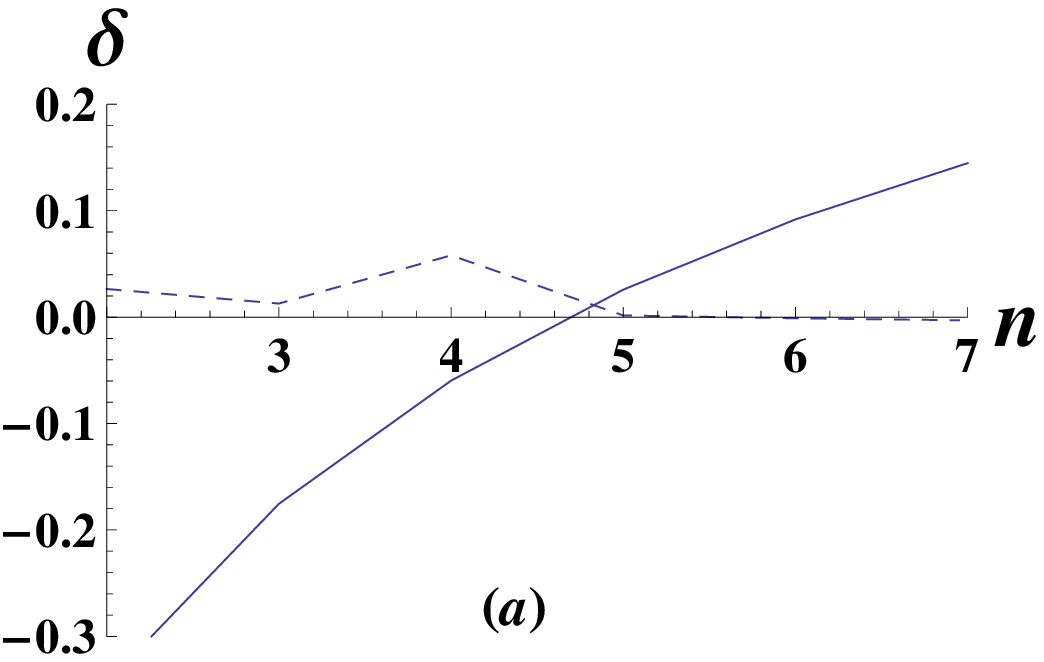}}
\put(16.5,0){\includegraphics{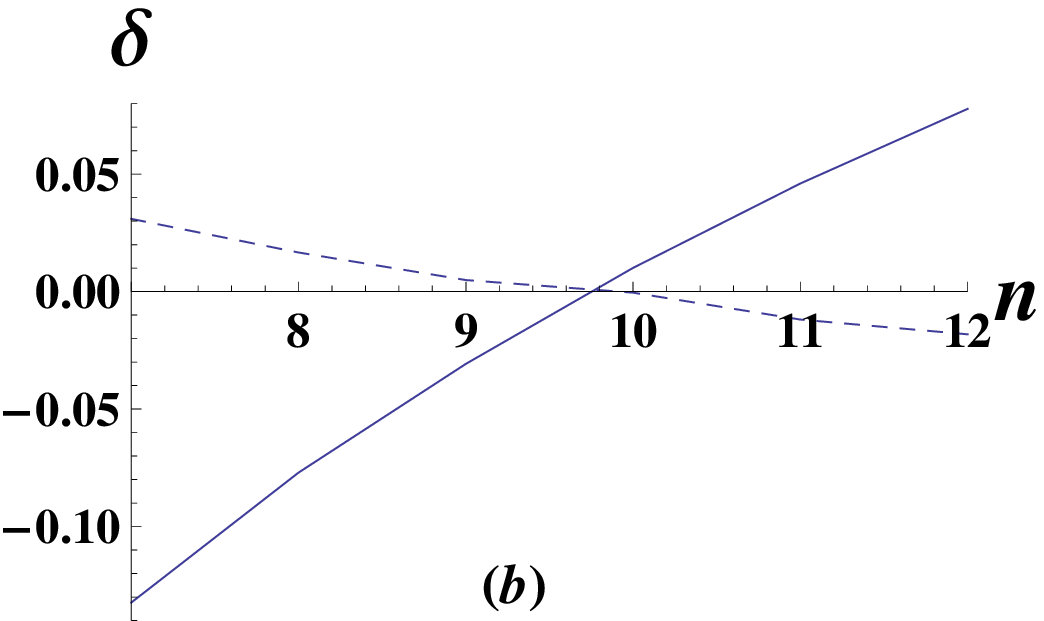}}
\end{picture}
\caption{ Comparison of 0-th order accuracy expression and the 1/L
accuracy expression for the $p_n$ in diploid model, for the
fitness function $f_0(m_1,m_2)=\frac{a}{2}(m_1^2+m_2^2)+ bm_1m_2$
and $a=4.5,b=0.5$.
 $p_n$ are
the results of numerics for the of the original model given by the
system of equations. $p_{0n}=\exp(Lu(m))$ is calculated using
$O(L)$ terms in expression of $\ln p_n$, Eq.(12), while
$p_{1n}=\exp(Lu(m)+u_1(m))$ corresponds to the $O(1)$ accuracy in
expression of $\ln p_n$, Eq. (12). $\delta=\ln p_{0n}/p_n$, the
smooth line. For Fig 2.a $L=50$ $\delta=\ln p_{1n}/p_n$ is the
dashed line. For Fig 2.b  $L=100$, and $\delta=10\ln p_{2n}/p_n$
is the dashed line.}
\end{figure}

\section{Horizontal Gene Transfer model}

The finite genome length effects are very important for the case
of horizontal gene transfer(HGT).  While the genome length could
be large (about 10000 in case of HIV), it is possible to consider
only the variable part of genome, $L=40-100$ \cite{bo05}). We
consider the model \cite{le05,de07} describing a simple
horizontal gene transfer. The model has been solved \cite{de07} in
the large genome limit in \cite{de07} (mean fitness) and
\cite{sa08} (steady state distribution).

In this model, besides the mutation process,  one spin in the
genome is replaced  with the spin at the same position from the
sequence pool.

We consider the following system of equations \cite{le05,de07}:
\begin{eqnarray}
\label{e33}
&&\frac{dP_l}{dt}= P_lr_l+[(L-l+1)P_{l-1}+(l+1)P_{l+1}]- \nonumber\\
 && P_l(1+\frac{1}{L}\sum_{k}r_kP_k)- \nonumber \\
 &&  cP_{l}+c(\frac{\bar l}{L}
\frac{l}{L} +
\frac{{L- l}}{L}\frac{L-\bar l}{L})P_{l}+ \nonumber\\
&&c(1-\frac{{l}-1}{L})\frac{\bar l-1}{L}P_{l-1}+
c\frac{{l}+1}{L}(1-\frac{\bar l+1}{L})P_{l+1}],
\end{eqnarray}
where $\hat l=\sum_lP_ll=(1-s)L/2$. The first and second lines
correspond to the Crow-Kimura model \cite{ck70},\cite{ba01}. The
third line describes  the HGT process with the change of the
Hamming class. We cut plus spin with the probability $(1-l/L)$ and
add plus spin from the sequence pool with the probability $(1-\bar
l/L)$, then cut minus spin with probability $l/L$ and add minus
spin from the sequence pool with the probability $\bar l/L$. The
last line corresponds to the  process with the change of  Hamming
class. We cut the plus spin with the probability $(l-1)/L$ and add
minus spin from sequence pool with the probability $1-(\bar
l-1)/L$, then cut the minus spin with the probability $(l+1)/L$
and add plus spin from sequence pool with the plus spin with the
probability $1-(\bar l+1)/L$.

Considering the ansatz $$P_l=\exp[L(k+\frac{k_1}{L})t+
Lu(m)+u_1(m)],$$ we get the following equation for the bulk terms
\begin{eqnarray}
\label{e34}
&& k=H(m,u')\nonumber\\
&& H(m,p)\equiv f(m)+\frac{1+m}{2}e^{2p}[1+\frac{c(1-s)}{2}]\nonumber\\
&+&\frac{1-m}{2}e^{-2p}[1+c\frac{1+s}{2}]
+\frac{cms}{2}-\frac{c}{2}-1,
\end{eqnarray}
where $r_l=Lf(m)$ and $p$ is a dummy variable. The minimum of
$H(m,u')$ is at $u'=u_0(m)$, where
\begin{eqnarray}
\label{e35} e^{4u_0(m)}=\frac{(1-m)(1+c\frac{1+s}{2})}
{(1+m)(1+c\frac{1-s}{2})}
\end{eqnarray}
gives a potential $V(m,s)\equiv min[H(m,u')]_{u'}$:
\begin{eqnarray}
\label{e36}
 &&V(m,s)=f(m)+\sqrt{(1-m^2)C}+\frac{cms}{2}-\frac{c}{2}-1,\nonumber\\
 &&C=[(1+\frac{c}{2})^2-\frac{c^2s^2}{4}].
\end{eqnarray}
We define $m_0,s$ from the conditions
\begin{eqnarray}
\label{e37}
 V'(m_0,s)=0,\quad f(s)=V(m_0,s),
\end{eqnarray}
where the derivative is with respect to  the first argument. Let
us calculate the $u''(m_0)$. Near $m_0$ we have an expansion
\begin{eqnarray}
\label{e38} &0\approx& V''(m_0,s)\frac{(m-m_0)^2}{2} \nonumber \\
&+&H''_{pp}(m_0,u'_0)\frac{(u'(m)-u_0(m))^2}{2},
\end{eqnarray}
where $V''$ is the second derivative of $V$ with respect to the
first argument, $H'_{pp}$ is the second partial derivative of $H'$
 with respect to the
second argument of the function $H(m,p)$. Dividing Eq.(\ref{e38})
by $(m-m_0)^2$, we derive the equation
\begin{eqnarray}
\label{e39}
 f''(m_0)=\frac{\sqrt{C}}{(1-m_0^2)^{3/2}}-\sqrt{(1-m^2)C}(u''-u''_0)^2
\end{eqnarray}
or
\begin{eqnarray}
\label{e40} 2u''(m_0)=-\frac{1}{(1-m_0^2)}- \frac{
\sqrt{-f''+\frac{\sqrt{C}}{(1-m_0^2)^{3/2}}}}{[(1-m_0^2)C]^{1/4}}.
\end{eqnarray}
From Eq.(36) we have:
\begin{eqnarray}
\label{e41}
 k_1
 &=&\frac{f'(s)u_1'(s)}{|u''|}+\frac{f''(s)}{2u''(s)}  \nonumber\\
 &=& 2u_1'[e^{2u'}\frac{1+m}{2}(1+c\frac{1-s}{2}) \nonumber \\
 &&-\frac{1-m}{2}(1+c\frac{1+s}{2})e^{-2u'}]\nonumber\\
&+& [e^{2u'}(1+c\frac{1-s}{2})+(1+c\frac{1+s}{2})e^{-2u'}] \nonumber\\
&+& 2u''[e^{2u'}\frac{1+m}{2}(1+c\frac{1-s}{2}) \nonumber \\
&&+ \frac{1-m}{2}(1+c\frac{1+s}{2})e^{-2u'}]\nonumber\\
&+&
cu_1'(s)\frac{m-e^{2u'}\frac{1+m}{2}+e^{-2u'}\frac{1-m}{2}}{2|u''|}.
\end{eqnarray}
The first line is a direct integration of the fitness via steady
state distribution, the second line corresponds to the correction of
$F(m,p)$ via $\delta p\equiv u_1'$, the third line corresponds to
the corrections in the first line of Eq.(\ref{e33}).
 The last line corresponds to the $\delta s$ corrections
 from the $\hat l$ terms in Eq.(\ref{e33}).

 To calculate $k_1$, we put the optimal value of $u'$ and look at the point
 $m=m_0$.
 Then Eq.(44) is simplified:
\begin{eqnarray}
\label{e42}&& \frac{f'(s)u_1'(s)}{|u''|}+\frac{f''(s)}{2|u''(s)|}
\nonumber \\
&=& \frac{2}{\sqrt{1-m_0^2}}\sqrt{C}+
2u''(m_0)\sqrt{1-m_0^2}\sqrt{C} \nonumber \\
&&+cu_1'(s) \frac{m_0-\frac{cs\sqrt{1-m_0^2}}{2{\sqrt{C}}} }
{2|u''(s)|}\nonumber\\
&=& cu_1'(s)
\frac{m_0-\frac{cs\sqrt{1-m_0^2}}{2{\sqrt{C}}} } {2|u''(s)|} \nonumber\\
&+&\frac{\sqrt{C}}{\sqrt{1-m_0^2}}
[1-\sqrt{1-\frac{f''(1-m_0^2)^{3/2}}{\sqrt{C}}}].
\end{eqnarray}
Deriving $u_1'(s)$ from the last equation, we obtain
\begin{eqnarray}
\label{e43}
 k_1
 =\frac{f'(s)u_1'(s)}{|u''|}+\frac{f''(s)}{2|u''(s)|}.
\end{eqnarray}
Thus
\begin{eqnarray}
\label{e44} k_1= \frac{f''(s)s}{f'(s)}+ \frac{f'(s)}{f'(s)-
c\frac{m_0-\frac{cs\sqrt{1-m_0^2}}{2{\sqrt{C}}} } {2}}\times
\end{eqnarray}
$$[\frac{\sqrt{C}}{\sqrt{1-m_0^2}}
[1-\sqrt{1-\frac{f''(m_0)(1-m_0^2)^{3/2}}{\sqrt{C}}}]-\frac{f''(s)}{f'(s)}]$$

For the fitness function $f(m)=cm^2$ with parameters $c=1$, we
derive from the numerics for the mean fitness and order parameters
$R/L=0.218142,m_0=0.765984,s=0.467057$. The analytical formula
Eq.(\ref{e37}) for the infinite genome limit gives $k=0.21842$,
while the numerics for $L=100$ gives $R/L=0.21973$ and for
$L=200,$ $R/L=0.218946$. Thus we get for $L=100$, $R/L-k=0.00158,
R/L-k-k_1/L=0.00002$ and for $L=200$, $R/L-k=0.000803,
R/L-k-k1/L=0.000003$. We see that our analytical result by Eq.
(\ref{e44}) for the mean fitness expression is well confirmed.

One can define $u_1(m)$ by integrating the expression of $u'_1(m)$
from Eq. (\ref{e42}).
\section{Discussion}

We calculated the finite genome size corrections to the mean
fitness and steady state distribution for strongly nonlinear
evolution models: horizontal gene transfer model  and diploid
parallel evolution model for general fitness function. The
application of the Hamilton-Jacobi equation for the investigation
of the master equation, especially the finite size corrections, is
a rather popular method in case of linear master equation for
chemistry or ecology \cite{es09},\cite{as10}. For the models,
considered in our article, the standard methods of linear algebra
and the methods of \cite{es09},\cite{as10} fail.

The key point of our method is that we investigate the point where
the coefficient of the correction to the steady state distribution
disappears. Then the correction to the fitness is calculated from
the smoothness condition. Our method could be applied to other
nonlinear probabilistic models as well.

 Our formulas could be
applied for relatively small genome lengths, for the single peak
fitness landscape  even for $L\sim 4$, see Appendix A. In
population genetics usually few allele models are investigated
with $L=2,3$. Fortunately, for the larger allele numbers we can
apply our methods. Our formulas, while a bit involved, are less
cumbersome than the formulas for several allele cases in
population genetics.

DBS thanks DARPA Prophecy Program and Academia Sinica for
financial support.


\renewcommand{\theequation}{A.\arabic{equation}}
\setcounter{equation}{0}

\section*{Appendix A. Finite $L$ corrections for
a parallel diploid  model with the single-peak fitness function}

Consider the following diploid model studied in \cite{sa08}:
\begin{eqnarray} \label{a1}
\frac{ dp_i}{dt}&=& p_i[\sum_jA_{ij}p_j
-\sum_{l=1}^{2^L}\sum_{j=1}^{2^L}A_{jl}p_jp_l] \nonumber \\
&+&\sum_jm_{ij}p_j
\end{eqnarray}
with the fitness coefficients
\begin{eqnarray}
\label{a2}
 A_{11}&=&2s;~
A_{1i}=A_{i1}=2sh,~{\rm for}~ i\ne 1;\nonumber\\
A_{ij}&=&0,~{\rm for}~i> 1,~j> 1.
\end{eqnarray}
Here $m_{ij}$ is the mutation matrix, $m_{ii}=-1,~ m_{ij}=1/L$ for
$d_{ij}=1$, and for other cases $m_{ij}=0$.

 For the steady state we have \cite{sa08,ba95}
\begin{eqnarray}
\label{a3}
R p_i&=& p_i\sum_jA_{ij}p_j
+\sum_jm_{ij}p_j,\nonumber\\
R&=&\sum_{l=1}^{2^L}\sum_{j=1}^{2^N}A_{jl}p_jp_l \nonumber \\
&=&2s[(1-2h)x^2+2hx].
\end{eqnarray}
Here $x$ takes one of the solution $x_{\pm}$ defined below
\begin{eqnarray}
\label{a4}
x_{\pm} = \frac {-b \pm \sqrt{b^2-4ac}}{2a},
p_1=x\frac{1}{f_1-f_2},
\end{eqnarray}
where $a=1-2h$, $b=3h-1$, and $c=1/(2s)-h$. We have for the
marginal fitness $f_1,~f_2$ and the fitness $R$
\begin{eqnarray}
\label{a5}
f_1=2sx+2sh(1-x),
f_2=2shx\,   \nonumber\\
R=xf_1+(1-x)f_2
 =2s[(1-2h)x^2+2hx].
\end{eqnarray}
In terms of the marginal fitness, the original equation (A1) can
be written in the form
\begin{eqnarray}
 \label{a6}
\frac{d P_l}{d t} &=& (r_l -R)P_l + (L-l+1) P_{l-1} \nonumber\\
&+& (l+1) P_{l+1} - L P_l,
\end{eqnarray}
where $P_l$ are the class probabilities of the configurations at
the Hamming distance $l$ from the reference sequence, $r_0=Lf_1$,
and $r_l=Lf_2,l\ge 1$.

Let us denote $x=x_0+\delta,~R=R_0+k_1/L$. Then
\begin{eqnarray}
\label{a7}
k_1&=&2s[(1-2h)x+2h]\delta, \nonumber\\
f_1&=&2sx+2sh(1-x)+2s(1-h)\delta.
\end{eqnarray}

For the $p_1$, we have the equation
\begin{eqnarray}
\label{a8} x(R(x)+1-f_1(x))= \frac{{1}}{L}p_{1}.
\end{eqnarray}
Replacing on the right hand side and before the parenthesis on the
left hand side
$$x \sim (1-1/(f_1-f_2))$$ and
$$p_1 \sim(f_1-f_2-1)/(f_1-f_2)^2,$$
$$\bar l \sim 1/(f_1-f_2-1),$$
we find a perturbation  expression for $\delta$
\begin{eqnarray}
\label{a9} \delta=\frac{1}{L(R'-f_1')}\frac{1}{f_1-f_2} + O
(\frac{1}{L^2})
\end{eqnarray}
and for $k_1=R'(x)*\delta$
\begin{eqnarray}
\label{a10} k_1&=&\frac{1}{L(f_1-f_2)(1-\frac{f_1'}{R'})}
\nonumber
\\ &=& \frac{1}{L(f_1-f_2)[1-\frac{s-h}{2[(1-2h)x+h]}]}.
\end{eqnarray}
For $s=5.3,h=0.1,L=4$ we get $R_n=8.55$, while $R_0+k_1=8.49$,
thus we have $1\%$ accuracy.


\begin{thebibliography}{30}
\bibitem{note} The left-hand side of Eq. (13) in \cite{sa08}
should be divided by $N$.
\end{thebibliography}
\end{document}